\documentclass[%
 reprint,
superscriptaddress,
 amsmath,amssymb,
 aps,
floatfix,
pre
]{revtex4-1}

\usepackage{graphicx}% Include figure files
\usepackage{dcolumn}% Align table columns on decimal point
\usepackage{bm}% bold math
\usepackage{epstopdf}
\usepackage{subfigure}
\usepackage{color}
\usepackage{longtable}
\usepackage{ulem}

\usepackage{hyperref}% add hypertext capabilities
\DeclareMathOperator{\erf}{erf}
\newcommand{\revision}{\textcolor{black}}

\begin{document}

%\preprint{APS/123-QED}

\title{Statistics of the separation between sliding rigid rough surfaces:\\Simulations and extreme value theory approach}% Force line breaks with \\

\author{Nicolas Ponthus}
\affiliation{%
Univ Lyon, Ecole Centrale de Lyon, ENISE, ENTPE, CNRS, Laboratoire de Tribologie et Dynamique des Syst\`emes LTDS UMR5513, F-69134, Ecully, France}
\author{Julien Scheibert}
\affiliation{%
Univ Lyon, Ecole Centrale de Lyon, ENISE, ENTPE, CNRS, Laboratoire de Tribologie et Dynamique des Syst\`emes LTDS UMR5513, F-69134, Ecully, France}
\author{Kjetil Th\o gersen}
\affiliation{%
Physics of Geological Processes, The NJORD Centre, Department of Geosciences, University of Oslo, Norway}%
\author{Anders Malthe-S\o renssen}
\affiliation{%
Department of Physics, University of Oslo, 0316 Oslo, Norway}%
\author{Jo\"el Perret-Liaudet}
\affiliation{%
Univ Lyon, Ecole Centrale de Lyon, ENISE, ENTPE, CNRS, Laboratoire de Tribologie et Dynamique des Syst\`emes LTDS UMR5513, F-69134, Ecully, France}

\date{\today}

\begin{abstract}
When a rigid rough solid slides on a rigid rough surface, it experiences a \revision{random motion in the direction normal to the average contact plane}. Here, through simulations of the separation at single-point contact between self-affine topographies, we characterize the statistical and spectral properties of this normal motion. In particular, its \textit{rms} amplitude is much smaller than that of the equivalent roughness of the two topographies, and depends on the ratio of the slider's lateral size over a characteristic wavelength of the topography. In addition, due to the non-linearity of the sliding contact process, the normal motion's spectrum contains wavelengths smaller than the smallest wavelength present in the underlying topographies. We show that the statistical properties of the normal motion's amplitude are well captured by a simple analytic model based on the extreme value theory framework, extending its applicability to sliding-contact-related topics. 
\end{abstract}

\maketitle

%\tableofcontents

\section{\label{Intro}Introduction}

 The interfacial separation $d$ between the surfaces of two solids brought close to one another is central to many interfacial processes. Those include attractive forces when the distance is small but finite (Van der Waals \cite{hamaker_Vanderwaals}, electrostatic \cite{Jackson}, Casimir forces \cite{casimir_attraction}), repulsive elastic forces when the distance vanishes \cite{Johnson, persson_relation_2007}, heat transfer and non-contact friction \cite{volokitin_heat_transfer}, electric conductivity \cite{Plouraboue2000} and permeability \cite{Plouraboue2000,Talon2010}. The evaluation of $d$ becomes difficult when the typical separation becomes of the order of the surface roughness, because the separation is now a random variable of the position along the interface. In this case, $d$ often refers to the average separation, between the mean planes of the two rough surfaces.

\begin{figure}[!htbp]
\centering
\includegraphics[width=\columnwidth]{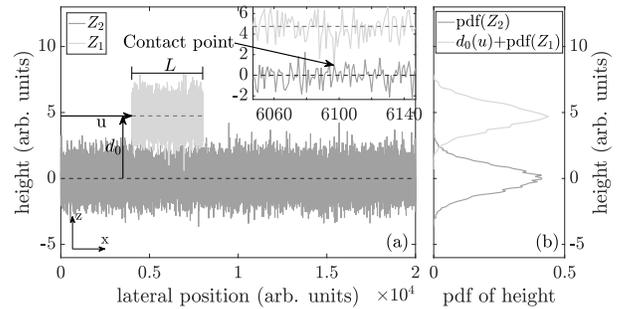}
\caption{\small{\revision{(a)}: Illustration of single-point contact, on the example of two 1D centered Gaussian white noises, $Z_1$ and $Z_2$. The separation at single-point contact, $d_0$, is measured between the mean heights of the two processes. \revision{(b)} Probability density functions (pdf) of both processes.}}
\label{fp_problem}
\end{figure}

In the particular case when the two rough surfaces come into contact, most of the literature has treated their normal approach (see \textit{e.g.} \cite{Vakis2018, Muser2017}, and \cite{Sahli2018} for shear loading). For elastic bodies under sufficient compressive pressure, a so-called multi-contact is formed, made of myriad individual micro-contacts where mainly the highest antagonist asperities are involved in the actual contact. This situation is typical of elastomer contacts \cite{Sahli2018}. The average separation between the two bodies is found to depend in particular on the ratio $p/E^\star$ of the applied pressure $p$ to composite elastic modulus $E^\star$ and the spectral properties of the topography \cite{persson_relation_2007, Yastrebov2017}. When $p/E^\star$ tends towards zero, \textit{i.e.} when the pressure becomes very low compared to the material stiffness, and when the two surfaces are brought in contact through a pure normal translation, those two surfaces touch on only one point, which is the first to come into contact. Such single-point contact situations, which are the focus of the present study, have previously been investigated in the context of the precise measurement of dispersion forces \cite{zwol_distance,broer_roughness} or of the contact of metallic surfaces under light load \cite{zouabi_these}. In such cases the measurable quantity is the separation of the two mean planes for single-point contact, $d_0$ (see Fig.~\ref{fp_problem}). Note that if the two solids are shifted one with respect to the other parallel to the contact plane, the measured value of $d_0$ will likely vary, because the single-point contact will involve a different couple of antagonist asperities. Such a sensitivity to details of the measurement procedure is responsible for significant uncertainties in the evaluation of $d_0$ \cite{zwol_distance,broer_roughness}. $d_0$ is also expected to vary as soon as the solids are slid one on another, because the point of contact will continuously change, and this is the phenomenon of interest in the following.

From now on, we will consider that, in such a weakly-loaded, single-point contact situation, one body (the slider, with a finite-sized area) is set to slide on the other (the track, having a larger area). Due to the random nature of the antagonist topographies, the slider will experience a roughness-induced motion in the direction normal to the \revision{average contact} plane \revision{($z$-displacement)}, $d_0(u)$, with $u$ the tangential displacement \revision{($x$-displacement)} of the slider, as shown on Fig.~\ref{fp_problem}. \revision{In the following, $d_0(u)$ is referred to as the normal motion.} If the sliding velocity was high enough, the slider could loose contact with the track and enter a bouncing regime (\cite{zouabi_these, Slavic, Dang_papier, zouabi_memory}). In the following, we only consider slow sliding, in which such inertia effects can be neglected. In those quasi-static conditions, the time dependence of the normal motion is irrelevant and the quantity of interest is $d_0(u)$. To characterize this quantity, we perform direct numerical simulations of the single-point contact between sliding rigid rough surfaces, as described in section~\ref{simu_method}.

From the illustration of Fig.~\ref{fp_problem}, it is natural to interpret the single-point contact sliding process as a geometrical filtering in which the input signals are the two antagonist topographies, $Z_1$ and $Z_2$, and the output signal is the roughness-induced normal motion, $d_0(u)$. In particular, one expects that the broader the probability density function of the topographies, the larger the average single-point contact separation. One also expects the spectral properties of $d_0(u)$ to be dependent on in-plane features of the topographies, like their spectral contents. This is why our simulations explore a variety of power spectrum densities (PSD) of the contacting topographies (section~\ref{simu_method}). In section \ref{simu_results}, we characterize in details the relationship between the properties of the topographies and that of the resulting normal motion $d_0(u)$.

The height of single-point contact being directly related to the altitude of the highest asperities of the antagonist surfaces, it is tempting to use the concepts of extreme value theory (EVT) \cite{Sornette,embrechts_modelling,de_haan_extreme} to estimate the statistical properties of $d_0$. EVT has been extensively used in various fields \cite{Fortin2015}, including rupture in disordered media \cite{Alava2006}, risk in finance or insurance \cite{embrechts_modelling}, or catastrophic natural events (preface of \cite{de_haan_extreme}). EVT predicts the probability distribution of rare events, and is used in section~\ref{Discussion} to predict the distribution of the maximum height of the topographies and thus of $d_0$. Those predictions are quantitatively compared to the simulation results and used to discuss the applicability of EVT to sliding-contact-related topics.

\section{\label{simu_method}Direct simulations: methods}
\subsection{\label{PropTopo}Properties of the topographies}
To characterize the properties of the separation at single-point contact between sliding surfaces, $d_0(u)$, we performed direct simulations of a rough square slider (surface $L \times L$) moving quasi-statically along a rough track (surface $L_1 \times L$, $L_1>2L$), and touching it in a single point for each of the successive positions $u$ of the slider. The two rotations of the slider around the \revision{in-plane} axis are forbidden and its translation along the track is imposed. Its only free motion is that along the \revision{out-of-plane, $z$-}axis. The slider and track have the same statistical roughness properties.

Two-dimensional (2D) Gaussian topographies, $\mathbf{z}$, with various spectral properties have been generated, from their 2D power spectrum density (PSD). Assuming that the topographies are isotropic, they are fully characterized by the radial profile of their PSD, $\mathcal{S}_{\mathbf{z}\mathbf{z}}(k_r)$, with $k_r$ the radial wave number. Knowledge of the PSD allows one to calculate a variety of useful estimators of the topographies' properties, among which its \textit{rms} roughness, $R_q$, from:
\begin{equation}
R_q^2=M_0,
\label{Eq:Rq}
\end{equation}
and its central wavelength, $\lambda_0$, from:
\begin{equation}
\lambda_0=\frac{1}{2}\sqrt{\frac{M_0}{M_2}},
\label{l_corr_eq}
\end{equation}
with the radial spectral moments $M_i$ defined by \cite{Longuet-Higgins} $M_i=2\pi\int_{0}^{+\infty}k_r^{i+1} \mathcal{S}_{\mathbf{z}\mathbf{z}}(k_r)\mathrm{d}k_r$.

\begin{figure}[!htbp]
\centering
\includegraphics[width=\columnwidth]{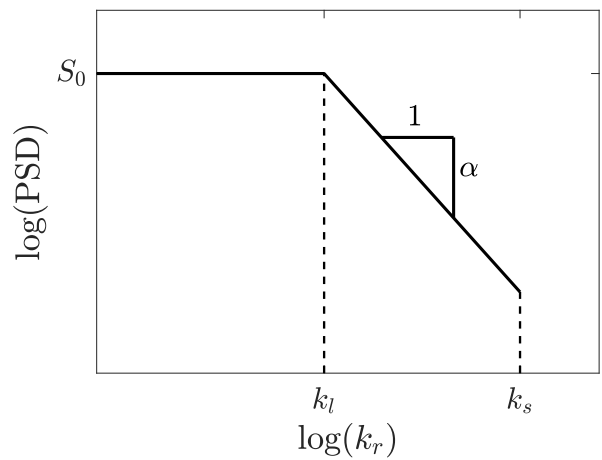}
\caption{Sketch of the radial profiles $\mathcal{S}_{\mathbf{z}\mathbf{z}}(k_r)$ of the 2D PSDs considered for the antagonist topographies and definition of the corresponding parameters.}
\label{schemaDSP}
\end{figure}

We used realistic PSDs corresponding to self-similar topographies such as the one shown on Fig.~\ref{schemaDSP}. Such PSD can be fully described using 4 parameters: $\mathcal{S}_0$ sets the amplitude of the surface; $k_{l}$, the low cut-off wave number, and $k_{s}$, the high cut-off wave number, set the wave number range over which the topographies are self-similar; $-\alpha$ is the slope of the self-similar part and is linked to the fractal dimension. Indeed, $\alpha$ relates to the Hurst exponent $H$ through \cite{long_noise,pal_frac_corr,Persson}: $\alpha=2(H+1)$. Note that, for a slider of size $L$, the lowest accessible wave number is $k_L=\frac{2\pi}{L}$. With this choice of PSD profile, the three first radial spectral moments $M_i$ can be calculated analytically (appendix \ref{surf_param_th}). Once injected in Eqs.~\ref{Eq:Rq} and~\ref{l_corr_eq}, appendix \ref{surf_param_th} provides the explicit expressions of $R_q$ and $\lambda_0$.

In summary, a given simulation corresponds to a given set of 5 parameters: $\mathcal{S}_0$, $k_{l}$, $k_{s}$, $\alpha$ and $L$. In practice, we will use the following equivalent set of 5 parameters with a more intuitive physical meaning: $R_q$, $\lambda_0$ and $L$ as three characteristic length scales and $\alpha$ and $b=\frac{k_l}{k_s}$ as two shape descriptors of the PSD radial profile.

\subsection{Numerical topography generation}

The surfaces are represented numerically by a $(2\Theta+1)\times (2\Psi+1)$ height matrix $\mathbf{z}$. The location along a surface is identified by the vector $\mathbf{x_{ij}}=(x_i,y_j)$ with $i$ varying from $-\Theta$ to $\Theta$ and $j$ from $-\Psi$ to $\Psi$ such that $x_i=i\Delta x$ and $y_j=j\Delta y$. Thus $\mathbf{z}(\mathbf{x_{ij}})$ defines the altitude of the topography at each point. The Fourier transform of $\mathbf{z}$ is $\mathcal{F}[\mathbf{z}](\mathbf{k_{\theta \psi}})=A(\mathbf{k_{\theta \psi}}) \exp(\mathrm{i}\phi(\mathbf{k_{\theta \psi}}))$ with $A$, the amplitude, and $\phi$, the phase, two real functions. The wave vector is $\mathbf{k_{\theta \psi}}=(k_{\theta}=\frac{\theta}{(2\Theta+1)\Delta x},k_{\psi}=\frac{\psi}{(2\Psi+1)\Delta y})$, with $\theta$ varying from $-\Theta$ to $\Theta$ and $\psi$ from $-\Psi$ to $\Psi$. By setting:
\begin{align}
A(n,m)&=A(-n,-m),\\
\phi(n,m)&=-\phi(-n,-m),
\end{align}
we ensure that $\mathbf{z}$ is real, and reads (after inverse Fourier transform of $A(\mathbf{k_{\theta \psi}}) \exp(\mathrm{i}\phi(\mathbf{k_{\theta \psi}}))$):
\begin{multline}
\mathbf{z}(\mathbf{x_{ij}})=\frac{1}{2\Theta+1}\frac{1}{2\Psi+1}\\
\sum_{\theta} \sum_{\psi} A(\mathbf{k_{\theta \psi}})\cos(\phi(\mathbf{k_{\theta \psi}})+\mathbf{k_{\theta \psi}}\cdot\mathbf{x_{ij}})
\label{Eq:znum}
\end{multline}
The amplitude $A(\mathbf{k_{\theta \psi}})$ can be expressed as a function of the continuous PSD profile, $\mathcal{S}_{\mathbf{z}\mathbf{z}}(k_r)$, as:
\begin{equation}
A(\mathbf{k_{\theta \psi}})=\sqrt{\frac{(2\Theta+1)(2\Psi+1)\mathcal{S}_{\mathbf{z}\mathbf{z}}(\left|\mathbf{k_{\theta \psi}}\right|)}{\Delta x\Delta y}}\label{Eq:A}.
\end{equation}

In order to produce numerical topographies obeying the PSDs described in section~\ref{PropTopo}, we use Eq.~\ref{Eq:znum} in which we insert both Eq.~\ref{Eq:A} and phases $\phi$ randomly drawn from a uniform law over $[0~2\pi[$, yielding a Gaussian distribution of heights.

\begin{figure}[!htbp]
\centering
\includegraphics[width=\columnwidth]{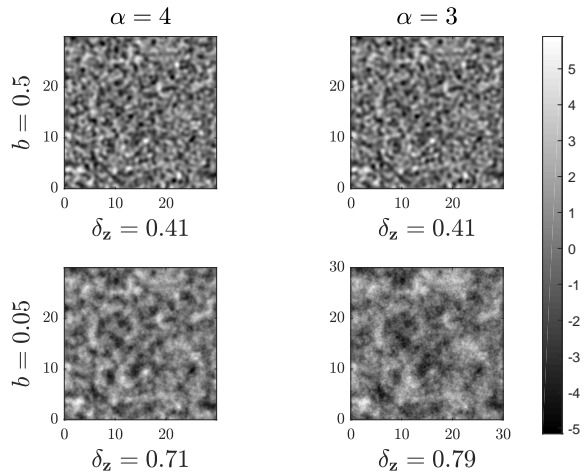}
\caption{Typical topographies generated, for various values of $\alpha$ and $b$. \revision{In-plane} size in units of $\lambda_0$. \revision{Out-of-plane} size (grayscale bar) in units of $R_q$.}
\label{exsurf}
\end{figure}

\begin{figure}[!htbp]
\centering
\includegraphics[width=\columnwidth]{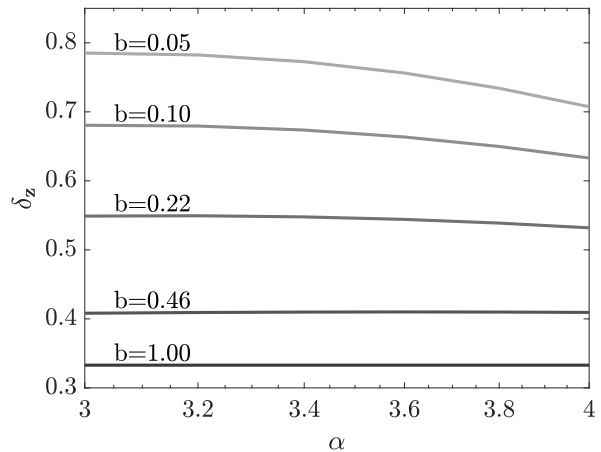}
\caption{Spectral bandwidth $\delta_\mathbf{z}$ as a function of the shape-descriptors of the PSD considered here (Fig.~\ref{schemaDSP}), $\alpha$ and $b$.}
\label{delta}
\end{figure}

Figure~\ref{exsurf} represents four typical topographies obtained for various values of $\alpha$ and $b$, the lateral length of all panels corresponding to the same number of central wavelength, $\lambda_0$. One can see that the smaller $b$ and $\alpha$, the richer the spectral contents of the topography, with $b$ having the strongest effect. The spectral bandwidth can be quantified by a spreading parameter $\delta_\mathbf{z}=\sqrt{1-\frac{M_1^2}{M_0M_2}}$ inspired by \cite{vanmarcke_properties,preumont,preumontbk,kiureghian,benasciutti}. $\delta_\mathbf{z}$ can vary between 0 and 1, with $\delta_\mathbf{z}$ being close to 0 for a narrow-band topography. Figure~\ref{delta}, which shows the evolution of $\delta$ as a function of $\alpha$, for various $b$, confirms the trends illustrated in Fig.~\ref{exsurf}.

\subsection{Simulation parameters}\label{simparam}

The in-plane discretization $\Delta x=\Delta y$ of the surface is chosen such that $\frac{2\pi}{2\Delta x}=6k_{s}$. This ensures that the sinus corresponding to the largest wave number is well-resolved, with twelve points per wavelength in the spatial domain. Sliding motion is simulated by moving the slider along $x$ with respect to the track, by one grid size at each step.

For each dimension of the topographies (\revision{out-of- and in-plane}) a reference length is chosen. For the \revision{out-of-plane} dimension, the \textit{rms} roughness, $R_q$, is chosen, while for the \revision{in-plane} dimension, we chose the central wavelength, $\lambda_0$. Note that, in our case of normal approach of rigid bodies, the in- and out-of-plane dimensions are uncoupled. In particular, dilating only the in-plane dimension does not affect the value of the normal separation, while dilating only the out-of-plane dimension does not affect the index of the topography points that are involved in the single-contact. Simulations are thus defined by three dimensionless parameters: two for the PSD shape, $b$ and $\alpha$, and one for the slider size $\tilde{L}=L/\lambda_0$. The output quantity is thus the dimensionless separation at single-point contact, $\tilde{d_0}=d_0/\sqrt{2}R_q$, as a function of the dimensionless sliding distance, $\tilde{u}=u/\lambda_0$. Note that, for the contact between two statistically identical topographies, each with an \textit{rms} roughness $R_q$, as is the case in the present study, the normalizing quantity for $d_0$, $\sqrt{2}R_q=R_q^*$, represents the equivalent \textit{rms} roughness of the sum topography.

The value of $b$ is varied from $0.05$ to $1$. Notice that $b=1$ is the case of a rectangular-shaped radial PSD, while for $b$=0, the topography would be purely self-similar. $\alpha$ is varied from $3$ to $4$, corresponding to a Hurst exponent varying from $0.5$ to $1$. To investigate the effect of the slider size, $\tilde{L}$ is varied from 43 to 760. The track size is then $\tilde{L}_1 \times \tilde{L}$. Note that we have limited the range of variations of $\tilde{L}$ to values such that (i) $k_L<k_l$ so that there is a white noise part in the PSD and (ii) $L_1=5L$ and $L/\Delta x$ is smaller than 17000, to keep topography matrices computationally tractable. Our computational resources allowed simulation of topographies with any combination of parameters within the above-mentioned ranges. An additional set of simulations has been performed with $\alpha=4$ and $b=0.46$, which allowed us to vary $\tilde{L}$ from 1.5 to 1389, and thus to explore more widely the effect of the slider size on the roughness-induced normal motion.

For each random draw of phases, a different topography is generated, with the specified PSD. For each set of parameters ($\alpha$, $b$ and $\tilde{L}$), several draws of topographies are performed in order to get converged statistical results for the separation at single-point contact, $\tilde{d_0}$. Tests have shown that with 15 draws, the expected value of each of the three first statistical moments (mean, standard deviation and skewness) of $\tilde{d_0}$  is measured to better than $5\%$ accuracy.

Finally, let us define the parameter $N$, which will be useful in the following sections:
\begin{equation}
N=\frac{4}{\pi}\tilde{L}^2.
\label{EqN}
\end{equation}
$N$ represents the number of circular patches of diameter $\lambda_0$ along the slider's surface. For a narrow band process, it is close to the number of asperities on the slider's surface. Here and in the following, the term asperity refers to any convex portion of the topography.

\section{\label{simu_results}Direct simulations: results}

\begin{figure}[!htbp]
\centering
\includegraphics[width=\columnwidth]{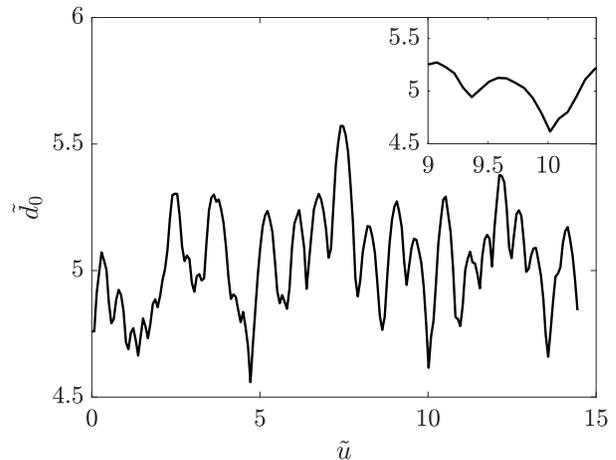}
\caption{Typical example of separation at single-point contact, $\tilde{d_0}(\tilde{u})$. $\alpha=4$, $b=0.46$, $\tilde{L}=327$. Inset: zoom showing cusp-like features.}
\label{ex_vib}
\end{figure}

\begin{figure}[!htbp]
\centering
\includegraphics[width=\columnwidth]{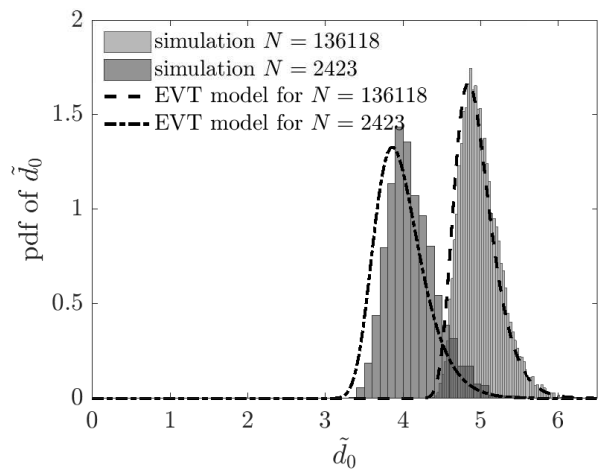}
\caption{Typical pdf of the separation at single-point contact for two different slider sizes, $\tilde{L}=327$ ($N=136118$) and $\tilde{L}=44$ ($N=2423$). $\alpha=3.2$, $b=0.46$. Dashed lines: EVT model discussed in section~\ref{Discussion}.}
\label{pdfnum}
\end{figure}

On Fig.~\ref{ex_vib}, an example of simulated separation at single-point contact, $\tilde{d_0}(\tilde{u})$, is plotted. On Fig.~\ref{pdfnum}, typical probability density functions (pdf) of $\tilde{d_0}$ are shown.
One can notice that $\langle \tilde{d}_0 \rangle$, the mean value of $\tilde{d}_0$, is larger than 0 by several $R_q^*$ (typically 2 to 5, depending of the simulated topogaphies). Note that $d_0(u)$ (the distance between the two mean planes (see Fig.~\ref{fp_problem})) can be equal to 0 only if one topography would be the exact complementary of the other at position $u$. Also, the standard deviation of $\tilde{d_0}$, $\sigma_ {\tilde{d}_0}$, is always found smaller than 1.
Finally, the skewness of the distribution, $sk_{\tilde{d}_0}$, is positive, due to the fatter right tail of the pdf, implying that, unlike the underlying topographies, the separation at single-point contact, $d_0$, is not a Gaussian process.

\begin{figure}[!htbp]
\centering
\subfigure{\label{moy}\includegraphics[width=0.95\columnwidth]{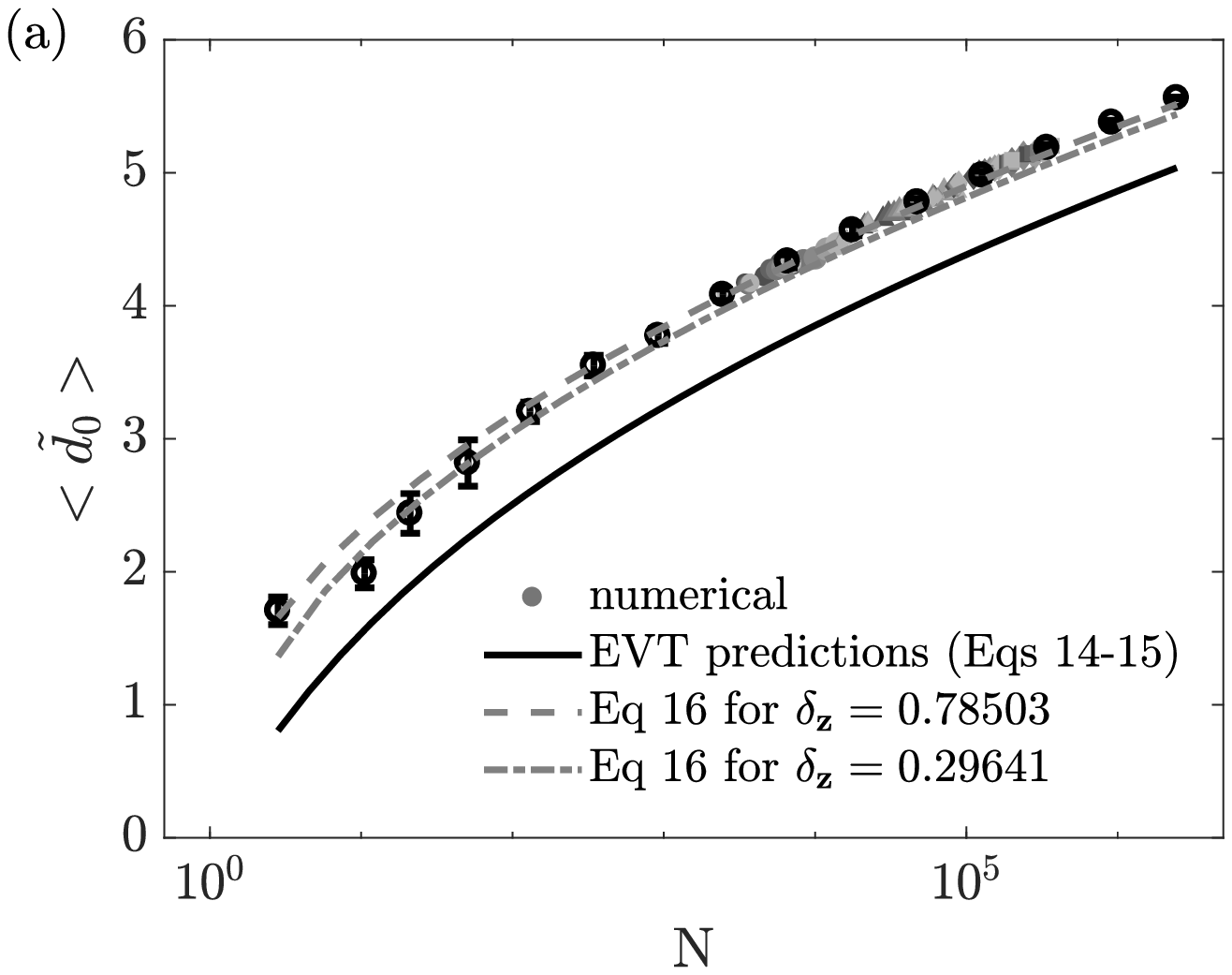}}
\subfigure{\label{std}\includegraphics[width=0.95\columnwidth]{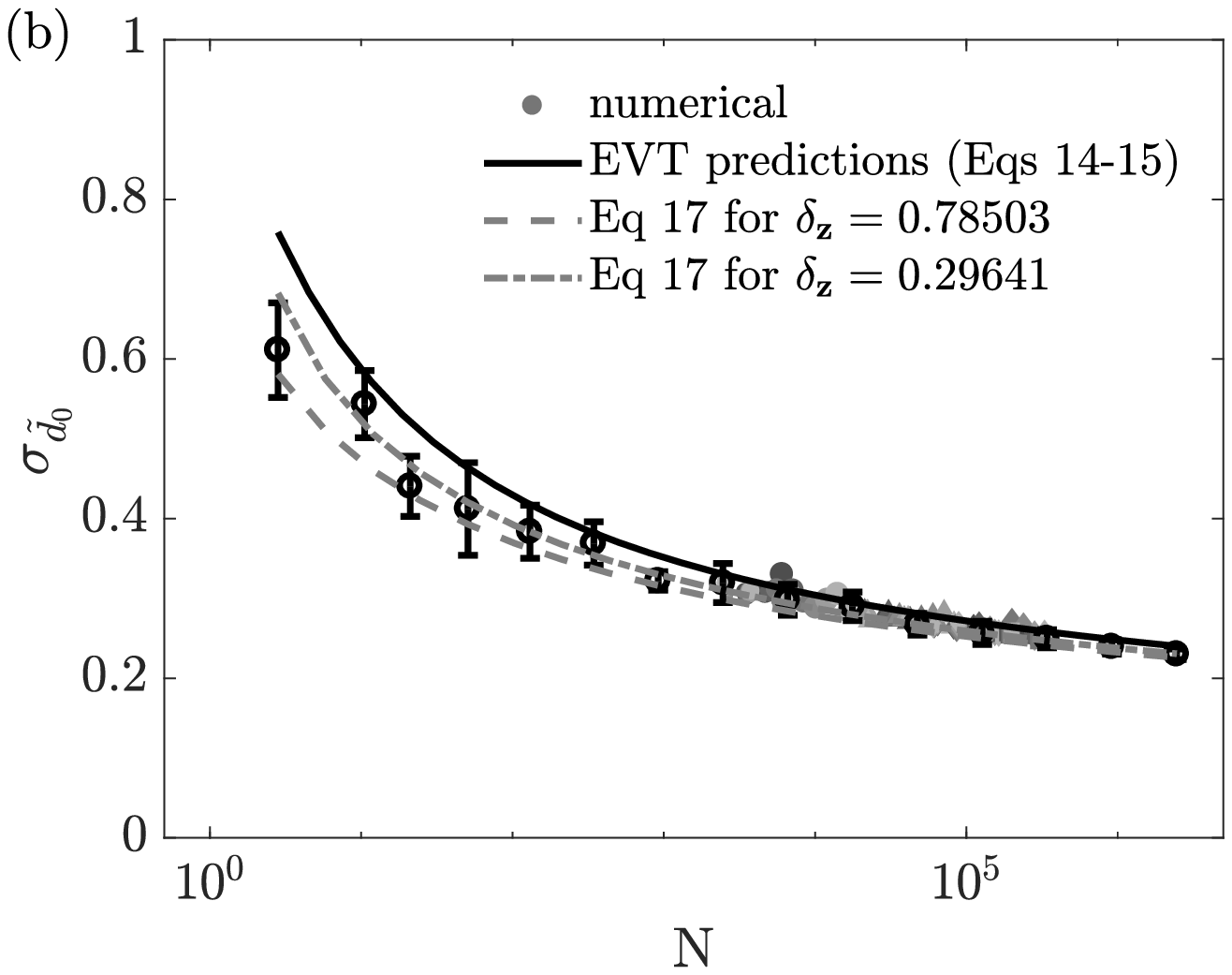}}
\subfigure{\label{ske}\includegraphics[width=0.95\columnwidth]{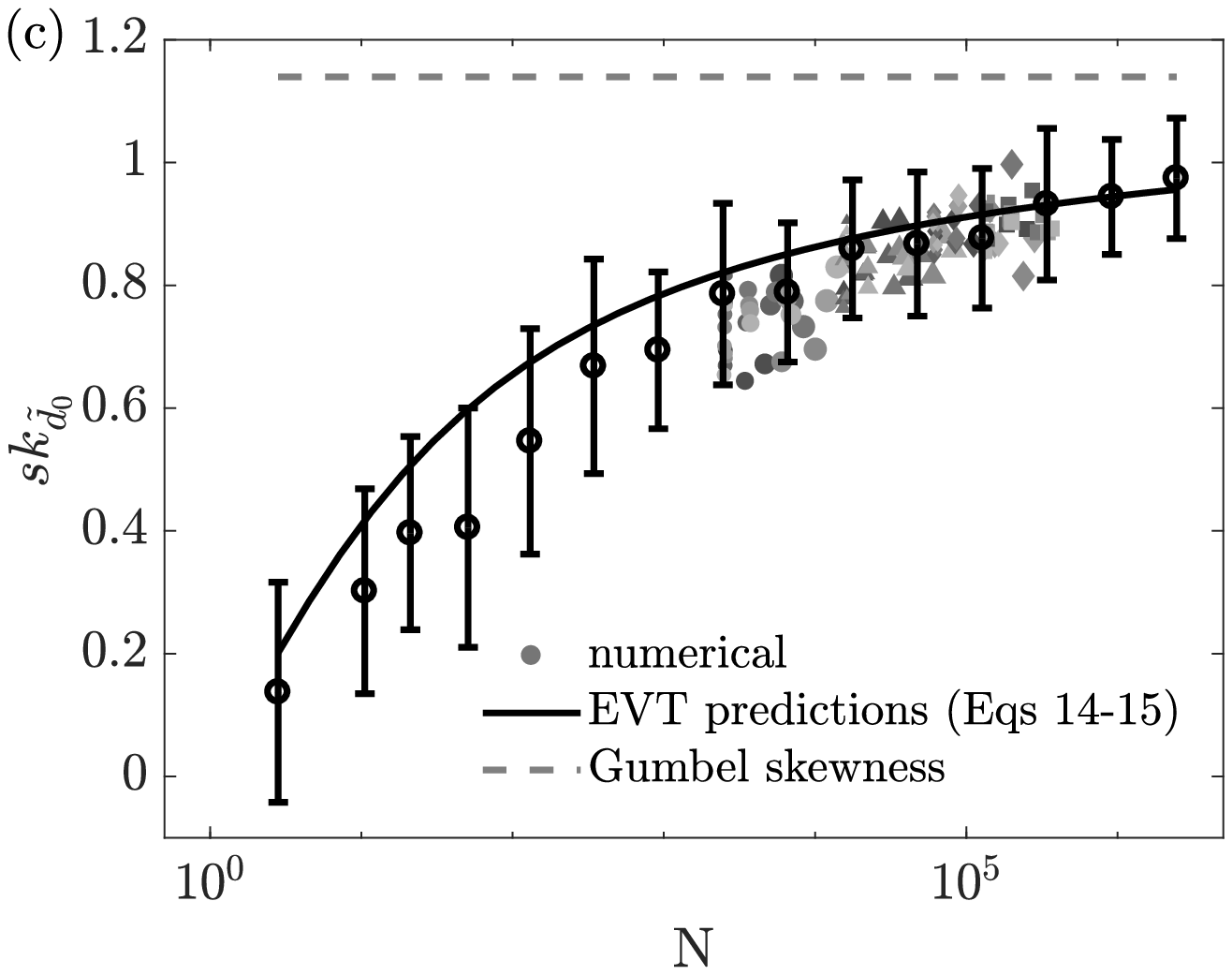}}
\caption{\revision{(a)} Mean value, \revision{(b)} standard deviation and \revision{(c)} skewness of the pdfs of $\tilde{d}_0$, for all simulations performed (circles). Error bars represent the standard deviation over 15 statistically identical simulations. For clarity, not all error bars are plotted. Lines: various models discussed in section~\ref{Discussion}. The marker size increases when $b$ decreases. Lighter gray is for  lower $\alpha$. $\bullet$, $\blacktriangle$, {\footnotesize$\blacklozenge$}, {\tiny$\blacksquare$} are for $L=100$, $250$, $500$ and $750$, respectively. The two values of $\delta_\mathbf{z}$ used in \revision{Figs.~\ref{moy} and \ref{std}} are the maximum and minimum values used in our simulations}
\label{momentum}
\end{figure}

More quantitatively, Fig.~\ref{momentum} shows the normalized mean value, standard deviation, and skewness, respectively $\langle \tilde{d}_0 \rangle$, $\sigma_ {\tilde{d}_0}$ and $sk_{\tilde{d}_0}$, for all simulations parameters used, as a function of the slider area, represented by the number $N$. Figure~\ref{momentum} clearly shows that the statistical moments of the dimensionless separation $\tilde{d_0}$ only depend on $N$. Both $\langle \tilde{d}_0 \rangle$ and $sk_{\tilde{d}_0}$ are increasing functions of $N$ whereas $\sigma_{\tilde{d}_0}$ is a decreasing function. Note that the $N$-axis is logarithmic, indicating that the variations with $N$ are relatively slow.

\begin{figure}[!htbp]
\centering
\includegraphics[width=\columnwidth]{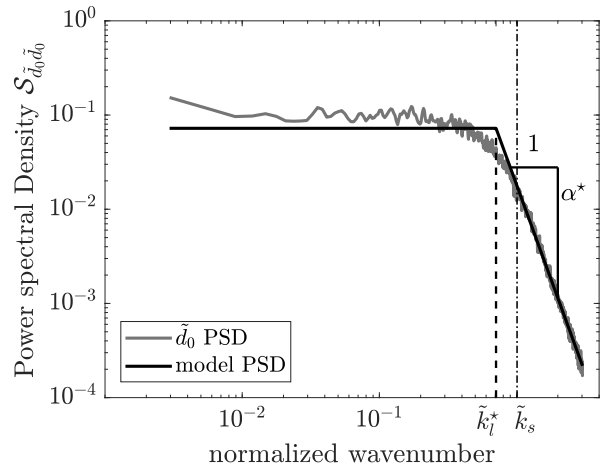}
\caption{Typical PSD of the separation at single-point contact and its empirical approximation (Eq.~\ref{k-4}). $\alpha=3.6$, $b=0.1$, $\tilde{L}=337$.}
\label{DSPnum}
\end{figure}

We now describe the spectral contents of the roughness-induced normal motion of the slider. They are described by the power spectrum density of $\tilde{d}_0(\tilde{u})$, $\mathcal{S}_{\tilde{d}_0\tilde{d}_0}$ (see Fig.~\ref{DSPnum} for a typical example). We find that the resulting PSDs are of the self-affine type, with a white noise part at low frequencies, and a power law decay of exponent $-\alpha^{\star}$ at high frequencies. The crossover wave number is denoted $\tilde{k}_l^{\star}$.

As can be seen on Fig.~\ref{DSPnum}, $\tilde{d_0}$ has non-vanishing spectral contents for wave numbers higher than $\tilde{k}_s$, the topographies' largest wave number. We measured the exponent $-\alpha^{\star}$ of the power-law decay of the PSD, for wave numbers larger than $\tilde{k}_s$, and found that it is always roughly equal to -4. Then, to estimate the value of $\tilde{k}_l^{\star}$, we propose the following empirical model for the PSD (see black line in Fig.~\ref{DSPnum}):
\begin{equation}\label{k-4}
\begin{cases}
\mathcal{S}_0^{\star} &	\text{if~}\tilde{k}<\tilde{k}_l^{\star},\\
\mathcal{S}_0^{\star}\left(\frac{\tilde{k}_l^{\star}}{\tilde{k}} \right)^4 &	\text{if~}\tilde{k}>\tilde{k}_l^{\star}.
\end{cases}
\end{equation}

\begin{figure}[!htbp]
\includegraphics[width=0.45\textwidth]{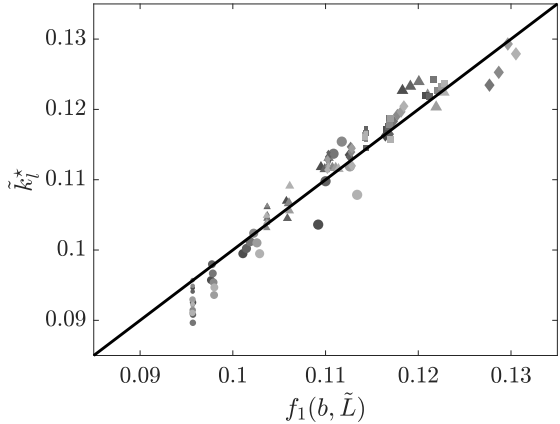}
\caption{Simulated $\tilde{k}_l^{\star}$ versus its approximated expression, $f_1(b,\tilde{L})$ (Eq.~\ref{Eqf1}). Solid line: equality line. The marker size increases when $b$ decreases. Lighter gray is for  lower $\alpha$. $\bullet$, $\blacktriangle$, {\footnotesize$\blacklozenge$}, {\tiny$\blacksquare$} are for $L=100$, $250$, $500$ and $750$, respectively.}
\label{spectralprop}
\end{figure}

We then fit the value $\tilde{k}_l^{\star}$ with the constraint that the moment of order 0 of the model PSD (which is the \textit{rms} value of the signal) is equal to that of the simulated one, which amounts to impose that $\mathcal{S}_0^{\star}=\frac{3}{4} \frac{\sqrt{\sigma_{\tilde{d}_0}^2+\langle \tilde{d}_0 \rangle^2}}{\tilde{k}_l^{\star}}$. Analysis of the dependence of $\tilde{k}_l^{\star}$ with the simulation parameters, $\alpha$, $b$ and $\tilde{L}$, for all the simulations performed, allowed us to find the following empirical expression:
\begin{equation}
\tilde{k}_l^{\star}\approx f_1(b,\tilde{L})=\left(\frac{5.18\cdot10^{-5}}{b^{1.54}}+0.0584\right)\tilde{L}^{0.0892}.\label{Eqf1}
\end{equation}
Figure \ref{spectralprop} shows that Eq.~\ref{Eqf1} nicely predicts the value of $\tilde{k}_l^{\star}$ obtained from the simulations. 

It is interesting to reformulate those results in terms of the central wavelength and the spectral bandwidth of the process $\tilde{d}_0$, which are generic estimators of a PSD, also valid in particular for non-self-affine PSDs. They are defined as $\tilde{\lambda}_0^{\star}=\frac{1}{2}\sqrt{\frac{\tilde{m}_0}{\tilde{m}_2}}$ and $\delta^{\star}=\sqrt{1-\frac{\tilde{m}_1^2}{\tilde{m}_0\tilde{m}_2}}$, respectively, with moments $\tilde{m}_i=\int_{-\infty}^{+\infty}{|\tilde{k}|^i\mathcal{S}_{\tilde{d}_0 \tilde{d}_0}(\tilde{k})\mathrm{d}\tilde{k}}$. Note that with such a definition~\cite{vanmarcke_properties,preumont,preumontbk,kiureghian,benasciutti}, odd moments do not vanish.

\begin{figure}[!htbp]
\subfigure{\label{delta_delta}\includegraphics[width=0.95\columnwidth]{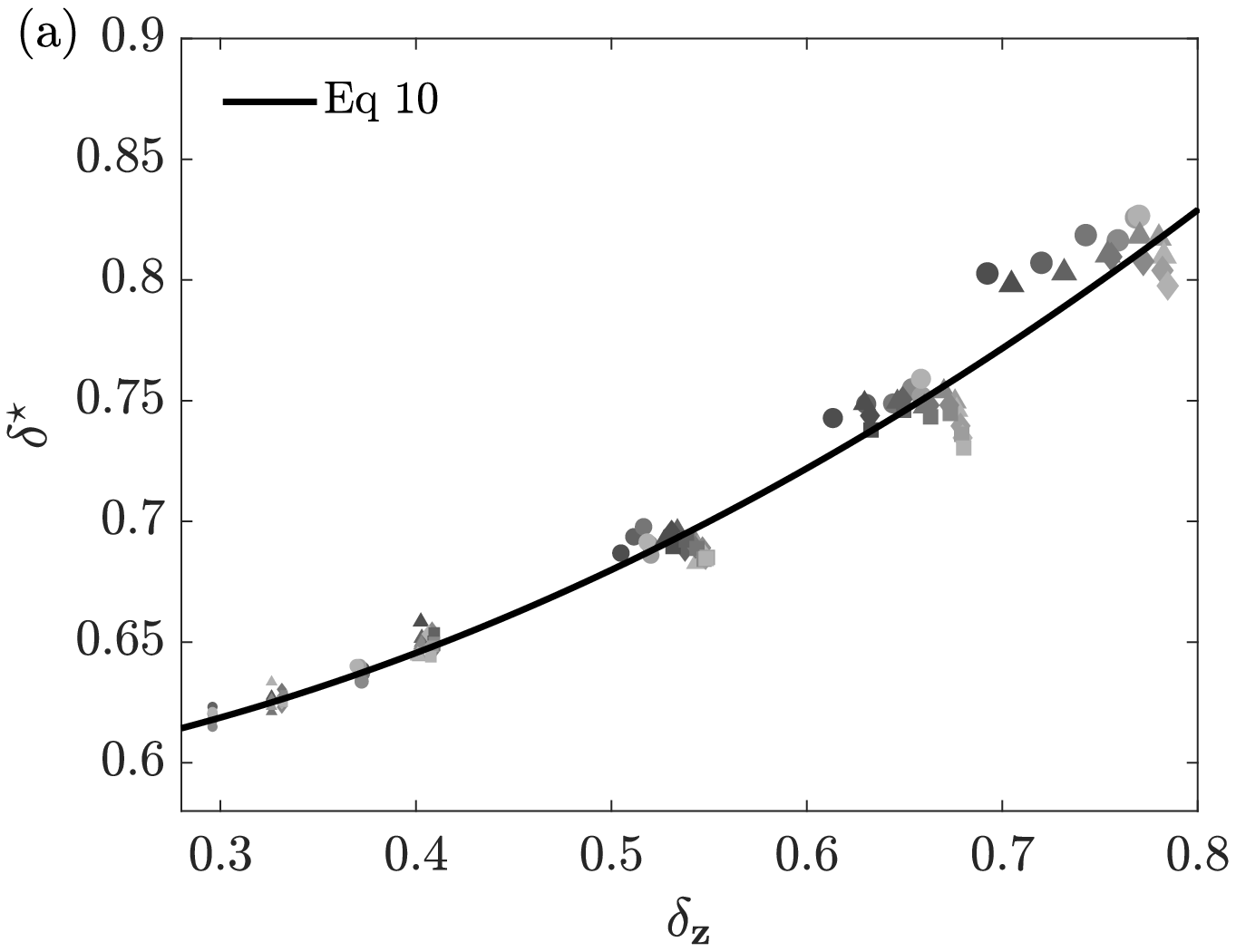}}
\subfigure{\label{lambda_tilde}\includegraphics[width=0.95\columnwidth]{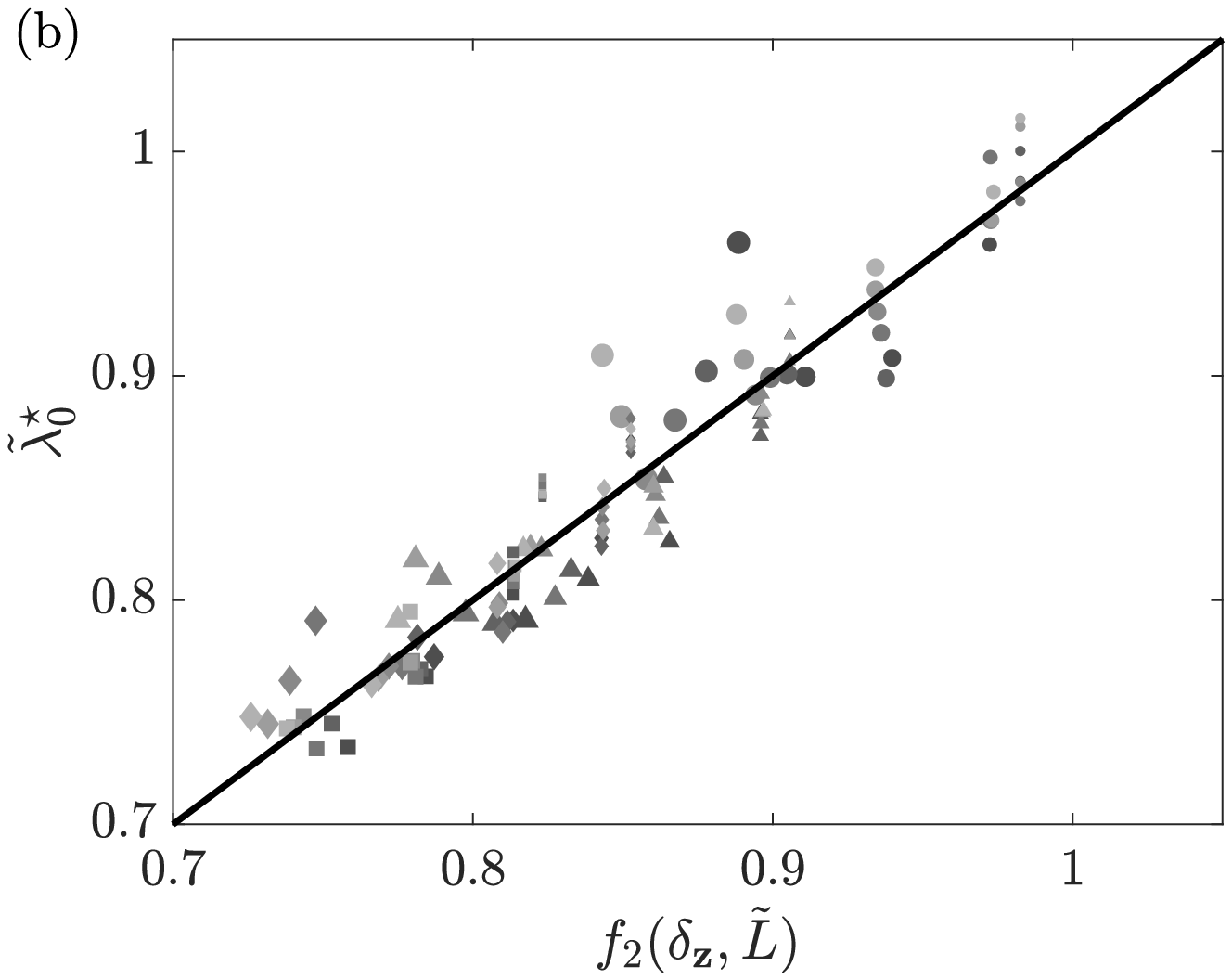}}
\caption{Spectral parameters of the normal motion $\tilde{d}_0(\tilde{u})$ as a function of those of the contacting surfaces. \revision{(a)} $\delta^{\star}$ vs $\delta_\mathbf{z}$. \revision{(b)} $\tilde{\lambda}_0^{\star}$ vs $f_2(\delta_\mathbf{z},\tilde{L})$ (Eq.~\ref{lambdatilde}). Solid line: equality line. The marker size increases when $b$ decreases. Lighter gray denotes lower $\alpha$. $\bullet$, $\blacktriangle$, {\footnotesize$\blacklozenge$}, {\tiny$\blacksquare$} are for $L=100$, $250$, $500$ and $750$, respectively.} 
\label{carac_band}
\end{figure}

We investigated the relationship between the spectral parameters of $\tilde{d}_0$ and those of the contacting topographies and found the results shown in Fig.~\ref{carac_band}. First, $\delta^{\star}$ is a function of $\delta_\mathbf{z}$ only, through (\revision{Fig.~\ref{delta_delta}}):
\begin{equation}
\delta^{\star}\approx0.38\delta_\mathbf{z}^2+0.58.\label{deltatilde}
\end{equation}

Second, $\tilde{\lambda}_0^{\star}$ depends on both $\delta_\mathbf{z}$ and $\tilde{L}$, through:
\begin{equation}
\tilde{\lambda}_0^{\star}\approx f_2(\delta_\mathbf{z},\tilde{L})=-0.178\delta_\mathbf{z}+\frac{1.50}{\tilde{L}^{0.0685}}.\label{lambdatilde}
\end{equation}
\revision{Figure~\ref{lambda_tilde}} shows that this expression is a good approximation of $\tilde{\lambda}_0^{\star}$, for all the simulations performed. Finding explanations for Eqs.~\ref{Eqf1},~\ref{deltatilde} and ~\ref{lambdatilde} would be the subject of an interesting future work.

\section{\label{Discussion}Discussion}
\subsection{\label{simulation_discussion}Single-point contact as a geometrical filtering}

The shape observed for the probability density function of the interfacial separation (Fig.~\ref{pdfnum}) can be understood as a geometrical filtering of the antagonist topographies. This filtering process is expected to strongly depend on the size of the slider. In the limit of a point-like slider ($L$ vanishes), it will be able to follow exactly the track's topography, so that its \revision{normal} motion will be equal to that topography. In the case of a finite-sized slider, the slider will not be able to penetrate into the valleys of the track's topography but will mainly slide on the highest asperities. Hence, the slider is expected to successively explore the shape of different asperities: the ones with the smallest distance to the slider's topography. The switching between asperities in contact is expected to be abrupt because the slider will intantaneously stop following the shape of the previous asperity and start following the new one. This scenario is in perfect agreement with the typical normal motion shown in Fig.~\ref{ex_vib} (inset), in which one can identify cusp-like features at the local minima (when the slider switches asperities) and smooth maxima (when the slider follows the summit of one asperity).

Those features of the normal motion are fully consistent with the observations made on the pdf and PSD of $\tilde{d}_0$. The asymmetry between minima and maxima in the normal motion explains the fatter tail of the pdf for large altitudes, which are more probable than the small amplitudes (at the cusps), and thus explains why $sk_{\tilde{d}_0}>0$ (\revision{Fig.~\ref{ske}}). It is interesting to note that the PSD of a cusp-containing signal like $|sin(x)|$ is a Dirac comb with amplitude decreasing asymptotically as $1/k^4$, \textit{i.e.} with an exponent close to the measured $-\alpha^{\star} \simeq -4$. We suggest that the presence of the cusps is the origin of the observed spectral enrichment (beyond $k_s$) of the roughness-induced normal motion. The observation that the standard deviation of $d_0$ is always smaller than that of the sum topography (\revision{Fig.~\ref{std}}) is related to the fact that the slider cannot explore the lowest parts of the track's topography, due to its finite size. Finally, the non-vanishing values of the mean of $\tilde{d}_0$ are fully consistent with the fact that the slider only touches the highest asperities of the track (\revision{Fig.~\ref{moy}}).

As noted above, the slider's normal motion only differs from the track's topography if the slider has a non-vanishing size, which indicates that the geometrical filtering is intrinsically a finite-size effect. And indeed, the various statistical properties of the normalized normal motion's pdf only depend on $N$ (Fig.~\ref{momentum}). Larger sliders have a larger probability to touch a high asperity, so that $\langle \tilde{d}_0 \rangle$ is larger (\revision{Fig.~\ref{moy}}). Similarly, larger sliders penetrate less into the tracks' valleys, so that $\sigma_{\tilde{d}_0}$ is smaller (\revision{Fig.~\ref{std}}).

\subsection{\label{EVT} Extreme value theory approach}

Due to the large number of points used to represent the topographies, the numerical simulations presented above are computationally expensive and require large Random Access Memory for the reverse Fourier transform operation (for the largest simulations, we used 512Gb of RAM for 3h30 on Bi-Xeon E5-2640v3 (16 core 2.6GHz)). Thus being able to predict the statistical properties of the separation at single-point contact, $d_0$, directly from the properties of the topographies is highly desirable. Remembering that the slider, due to its finite size, can only get into contact with the highest asperities of the track's topography, it is tempting to investigate how the framework of extreme value theory (EVT, see \textit{e.g.} \cite{Sornette,embrechts_modelling,de_haan_extreme}) can be applied to the present single-point contact problem.

We represent rough surfaces through $N$ points which are independent realizations of a centered Gaussian process. Consider two antagonist such topographies, $z_1$ and $z_2$, with identical \textit{rms} roughness $R_q$, as sketched in Fig.~\ref{fp_problem}. Their separation at single-point contact, $d_0$, is given by $d_0=-\min_{(x_i,y_j)}{(z_1(x_i,y_j)-z_2(x_i,y_j))}$. $z_1$ and $z_2$ having symmetric distributions, (i) the distribution of their difference is then statistically equal to the distribution of their sum and (ii) the opposite of the minimum is statistically equivalent to the maximum. We can thus work on the sum of the two topographies, $z=z_1+z_2$, which has a centered Gaussian distribution with a standard deviation equal to $R_q^*=\sqrt{2}R_q$, and examine $d_0=\max(z)$.

Let $\mathcal{P}$ be the cumulative density function (cdf) of $z$ and $p$ the associated probability density function (pdf). In our case:
\begin{align}
p(z)&=\frac{1}{R_q^*\sqrt{2\pi}}\exp\left(-\frac{z^2}{2{R_q^*}^2}\right),\\
\mathcal{P}(z)&=\frac{1}{2}\left(1+\erf\frac{z}{R_q^*\sqrt{2}}\right).
\end{align}
We will now follow the EVT approach described \textit{e.g.} in~\cite{Sornette}. The probability that the altitude at one point of $z$ is smaller than a value $Y$ is given by $\mathcal{P}(Y)$. The probability that all $N$ altitudes of $z$ are smaller than $Y$, meaning that $Y$ is greater or equal to the highest point of $z$, is the cdf $G(Y)=\mathcal{P}(Y)^N$. Thus, the pdf corresponding to the fact that $Y$ is the largest of the $N$ altitudes, which is precisely the pdf of the sliders' height, reads:
\begin{equation}
g(Y)=G'(Y)=Np(Y)\mathcal{P}(Y)^{N-1}.
\label{eq_pdf}
\end{equation}

Typical distributions $g$ are shown in dashed lines on Fig.~\ref{pdfnum} for two values of $N$, which is analogous to two different sizes of the slider. It appears that those EVT-predicted distributions present similar qualitative features as the simulated distributions. In particular, the mean height of the slider also increases with $N$, while the standard deviation of the slider's height also decreases with $N$. As in simulations, $g$ is not a symmetric distribution, but has a fatter tail for large values of $Y$ (positive skewness), which is typical of extreme value statistics. As shown in \cite{Sornette, leadbetter}, when $N$ is increased, the distribution of the maximum of a variable tends toward either the Frechet, Gumbel or Weibull distribution depending on the pdf of this variable. In particular, for a pdf with a right tail decaying faster than a power law, as is the case for Gaussian distributions, the pdf of the maximum will tend toward the Gumbel distribution. As a consequence, we expect the Gumbel distribution to be the limiting case of $g$ for very large sliders ($N\gg1$). Yet the convergence toward these limiting distributions is slow \cite{Majumdar} and thus they will not be reached here.

\subsection{\label{Comparaison} Comparison between methods}

Once the topography's pdf has been chosen to be Gaussian, the EVT prediction (Eq.~\ref{eq_pdf}) only depends on the parameter $N$. So, quantitative comparison between the predictions of EVT and the numerical simulations only relies on a relevant choice of $N$. Remembering that in EVT, we represent the topographies as a collection of $N$ discrete, statistically independent values, one looks for a number related to the number of asperities present on the simulated slider's surface. For a 1D process, this number can be given~\cite{preumont} by dividing the length of observation, $L$, by the central wavelength, $\lambda_0$ (Eq.~\ref{l_corr_eq}), so that $N=\frac{L}{\lambda_0}$. For the two-dimensional processes observed here, the same path of thought can be followed with areas of diameter $\lambda_0$:
\begin{equation}
N=\frac{L^2}{\pi\frac{\lambda_0^2}{4}}=\frac{4}{\pi}\left(\frac{L}{\lambda_0}\right)^2=\frac{4}{\pi}\tilde{L}^2,\label{EqN2}
\end{equation}
which justifies the prefactor used in Eq.~\ref{EqN}. Note that in \cite{zwol_distance}, a correlation length is used instead of $\lambda_0$ to define $N$.

Using this choice of $N$ in the EVT approach, we overplotted the analytical results of Eq.~\ref{eq_pdf} on all panels of Fig.~\ref{momentum}. This comparison shows a rather good quantitative agreement with our simulation results, confirming that the good match observed on Fig.~\ref{pdfnum} is actually true for all the explored simulation parameters. Yet, an offset exists on $\langle \tilde{d}_0 \rangle$ between the EVT prediction and the numerical results. We interpret this offset as a side effect of the approximation of a continuous topography by a set of discrete points: while the mean plane of a single, continuous asperity always lies below its summit, in the case of a point-like asperity, the mean plane has the exact same altitude as the summit itself. Hence, the mean planes of two continuous topographies in contact are always separated by a larger distance than those of two sets of discrete asperity summits. The exact offset between the two situations depends on both the amplitude and shape of the asperity. In our case, an empirical correction of $\frac{R_q^*}{2}$ seems to correctly capture our simulation data. Note that in Fig.~\ref{pdfnum}, the analytical pdfs shown include this correction, while the solid line in \revision{Fig.~\ref{moy}} does not.

We emphasize that such an agreement is \textit{a priori} non-trivial. First, the agreement quality significantly depends on the definition of $N$, suggesting that the arbitrary definition used (Eq.~\ref{EqN2}) is adequate. Second, the prediction is based on the EVT framework, which considers topographies made of independent realizations of a Gaussian process. In constrast, the topographies used in the simulations incorporate a finite correlation length, due to the shape of the PSD used to generate them. We believe that this difference is the main reason for the slight discrepancies observed in Fig.~\ref{momentum} between simulations and EVT predictions. To improve the agreement, one would need to account for the deviations from EVT induced by a finite correlation length.

This is what Preumont attempted in~\cite{preumont}, on the problem of finding the maximum value reached during a certain time window by a correlated 1D Gaussian signal. Assuming that the successive extrema of the signal form a Markovian process, he was able to find an exact, but intricate expression for the pdf of this maximum value.  By fitting this pdf with a Gumbel distribution, he was able to identify semi-empirical expressions of its mean value and standard deviation, as a function of $N$ and the spectral bandwidth $\delta$ of the process:
\begin{align}\label{preumont1}
{\langle \tilde{d}_0 \rangle}_{Preumont}&=\sqrt{2\ln \kappa_uN}+\frac{\gamma}{\sqrt{2\ln \kappa_\alpha N}},\\ 
{\sigma_{\tilde{d}_0}}_{Preumont}&=\frac{\pi}{\sqrt{6}}\frac{1}{\sqrt{2\ln \kappa_\alpha N}},\\
\kappa_u&=
\begin{cases}
1.5(1-e^{-1.8\delta}) 						&	\text{if~}\delta<0.5\\
0.94                                    &	\text{if~}\delta  \geq0.5, \label{preumont3}
\end{cases}\\
\kappa_\alpha&=
\begin{cases}
7\delta 						&	\text{if~}\delta<0.5\\
4.05                &	\text{if~}\delta  \geq0.5, \label{preumont4}
\end{cases}
\end{align}
with $\gamma=0.5772$ being Euler's constant. Note that the skewness of the Gumbel distribution is equal to $\frac{12\sqrt{6}\zeta(3)}{\pi^3}\sim1.14$ ($\zeta$ is Riemann's zeta function), independently of $N$ (see \revision{Fig.~\ref{ske}}).

Those semi-empirical expressions are overplotted on Fig.~\ref{momentum} using the value of $\delta_{\mathbf{z}}$ for $\delta$ in Eqs.~\ref{preumont3}~-~\ref{preumont4}. Those expressions appear to provide an excellent agreement with our simulation data, in particular they capture the correct amplitude of $\langle \tilde{d}_0 \rangle$. Such an improvement of the agreement confirms that the discrepancies observed between EVT and simulations are mainly due to the finite correlation of the simulated topographies. Yet, here again, such a good agreement was not expected, since Eqs.~\ref{preumont1}-~\ref{preumont4} were obtained for 1D processes, while our simulations use correlated 2D processes (the topographies).

\subsection{\label{Expe}Relation to experiments}

There are very few experimental works in the literature reporting measurements of the roughness-induced normal motion of macroscopic sliding solids. A notable exception can however be found in \cite{SoomKim1983a, SoomKim1983b}, where the authors monitor the normal acceleration of mild-steel slider-buttons of centimetric radius of curvature, during sliding on a rough mild steel disk. In particular, they report large wave number tails of the normal displacement PSDs of the type $k^{-4}$, in close agreement with our numerical findings (see Eq.~\ref{k-4}).

An interesting comparison can also be made with the literature about stylus measurements of rough surfaces. All wavelengths of the topography that are smaller than the tip size will be filtered-out through a geometrical filtering process analogous to the one studied here, leading to erroneous topography measurements (see \textit{e.g.} \cite{pastewka}). \revision{Indeed, in \cite{lechenault}, it is shown} that while the amplitude of large wavelengths is accurately measured, that of small wavelengths is underestimated. Thus, the \textit{rms} value of the measurement is smaller than that of the topography, consistently with our results of Fig.\ref{momentum}b. They also show that the crossover wavelength separating both regimes scales as $R^{1/(2-H)}$, with $R$ the curvature radius of a parabolic tip and $H$ the Hurst exponent. This result indicates a size-dependence of the filtering process, which is analogous to the $L$- (or $N$-) dependence that we observed. \revision{In \cite{church}, the authors} further showed that geometrical filtering induces cusps in stylus measurement, and that those cusps are responsible for a $k^{-4}$ behaviour of the large wave number tail of the PSD. Again, this is fully consistent with our results (see Eq.~\ref{k-4}).

%Building further on the qualitative analogy with stylus measurements suggested by the above-mentioned works, we can hypothesize, based on Fig. 7 (bottom), that a skewed stylus-measured height distribution may not only indicate a skewness in the topography itself, but also a possible effect of the finite-sized tip used to measure it.

\section{\label{Conclu}Conclusion}

We addressed the question of the roughness-induced normal motion during sliding of two solids in the limit of vanishing normal load, \textit{i.e.} when the contacting asperities do not deform. We considered the simplified case of the quasi-static evolution of the separation at single-point contact, when the slider has no rotational degree of freedom. Systematic numerical simulations assuming Gaussian self-affine topographies with various power spectrum densities, and sliders with various sizes have been performed. We found that the normal motion relates to the topographies through a geometrical filtering process which depends on the size of the slider. We also found that the resulting normal motion (i) has enriched spectral contents in the high wave number range, (ii) is non-Gaussian and (iii) has standard deviation much smaller than that of the sum topography. We provided empirical expressions relating the characteristics of the topography to that of the roughness-induced normal motion. We demonstrated that the distribution of the amplitude of the normal motion can be well predicted within the framework of extreme value theory (EVT) as soon as the number of points representing the topography of the slider is taken equal (to a prefactor close to 1) to the surface of the slider divided by the square of the central wavelength of the topography.

These results are relevant whenever rough surfaces are brought into light contact, that is, when there is no significant deformation of the bodies. They can be useful not only for sliding surfaces, but also to assess the variability of static measurements made on statistically equivalent contacts~\cite{zwol_distance,broer_roughness}. In particular, such a variability is expected to be much smaller than the characteristic amplitude of the two antagonist topographies. Our results are limited to single-point contacts, when the two solids are brought into contact through normal translation. In the case where a slider would be free to tilt, it would, under gravity, settle on three contact points to satisfy isostatic equilibrium. Accounting for such an effect is an interesting topic for a future work.

The fact that EVT nicely predicts the simulation results indicate that computationally expensive simulations like those decribed here may not be necessary in the future. Indeed, simple analytical formula (Eqs.~\ref{eq_pdf}-~\ref{EqN2}) or semi-empirical expressions (Eqs.~\ref{preumont1}-~\ref{preumont4}) are sufficient to evaluate most of the relevant statistical descriptors of the roughness-induced normal motion. Our results thus further extend the already large range of applicability of EVT to rough contact situations.

\begin{acknowledgments}
This work was supported by LABEX MANUTECH-SISE (ANR-10-LABX-0075) of Universit\'e de Lyon, within the program Investissements d'Avenir (ANR-11-IDEX-0007) operated by the French National Research Agency (ANR). It received funding from the People Program (Marie Curie Actions) of the European Union's Seventh Framework Program (FP7/2007-2013) under Research Executive Agency Grant Agreement PCIG-GA-2011-303871. J.P.-L. is member of the Labex CeLyA of Universit\'e de Lyon, operated by the French National Research Agency (ANR-10-LABX-0060/ANR-11-IDEX-0007). K.T. acknowledges support from EarthFlows - A strategic research initiative by The Faculty of Mathematics and Natural Sciences at the University of Oslo.
\end{acknowledgments}

\bibliographystyle{apsrev4-1}

\appendix

\section{Surfaces parameter}
\label{surf_param_th}
For surfaces described with radial power spectrum densities as shown in Fig.~\ref{schemaDSP}, the three first radial moments $M_0$, $M_1$, and $M_2$ have the following expressions:
\begin{align}
M_0&=2\pi\mathcal{S}_0\left(\frac{k_l^2-k_L^2}{2}-\frac{k_l^\alpha(k_s^{2-\alpha}-k_l^{2-\alpha})}{\alpha-2}\right)\\
M_1&=
\begin{cases}
2\pi\mathcal{S}_0\left(\frac{k_l^3-k_L^3}{3}-\frac{k_l^\alpha(k_s^{3-\alpha}-k_l^{3-\alpha})}{\alpha-3}\right) &	\text{if~}\alpha\ne3\\
2\pi\mathcal{S}_0\left(\frac{k_l^3-k_L^3}{3}+k_l^3(ln(k_s)-ln(k_l))\right) &	\text{if~}\alpha=3
\end{cases}\\
M_2&=
\begin{cases}
2\pi\mathcal{S}_0\left(\frac{k_l^4-k_L^4}{4}-\frac{k_l^\alpha(k_s^{4-\alpha}-k_l^{4-\alpha})}{\alpha-4}\right) &	\text{if~}\alpha\ne4\\
2\pi\mathcal{S}_0\left(\frac{k_l^4-k_L^4}{4}+k_l^4(ln(k_s)-ln(k_l))\right) &	\text{if~}\alpha=4
\end{cases}
\end{align}
\end{document}